\documentclass[aps,prl,twocolumn]{revtex4-1}

\usepackage{mathptmx}
\usepackage{graphicx}

\usepackage{CJKutf8} 
\usepackage{ucs} 
\usepackage[encapsulated]{CJK} 
\begin{document}
  
\title{Oscillate Boiling}
\author{Fenfang Li}
\affiliation{Division of Physics and Applied Physics, School of Physical and Mathematical Sciences, Nanyang Technological University, Singapore}
\author{S. Roberto Gonzalez-Avila}
\affiliation{Division of Physics and Applied Physics,
  School of Physical and Mathematical Sciences, Nanyang Technological
  University, Singapore}
\author{Dang Minh Nguyen}
\affiliation{Division of Physics and Applied Physics,
  School of Physical and Mathematical Sciences, Nanyang Technological
  University, Singapore}
\author{Claus-Dieter Ohl}
\email{cdohl@ntu.edu.sg}
\affiliation{Division of Physics and Applied Physics, School of Physical and Mathematical Sciences, Nanyang Technological University, Singapore}

\begin{abstract}
We report about an intriguing boiling regime occurring for small heaters embedded on the boundary in subcooled water. The microheater is realized by focusing a continuous wave laser beam to about $10\,\mu$m in diameter onto a 165\,nm-thick layer of gold, which is submerged in water. After an initial vaporous explosion a single bubble oscillates continuously and repeatably at several $100\,$kHz. The microbubble's oscillations are accompanied with bubble pinch-off leading to a stream of gaseous bubbles into the subcooled water. The self-driven bubble oscillation is explained with a thermally kicked oscillator caused by the non-spherical collapses and by surface pinning. Additionally, Marangoni stresses induce a recirculating streaming flow which transports cold liquid towards the microheater reducing diffusion of heat along the substrate and therefore stabilizing the phenomenon to many million cycles. We speculate that this oscillate boiling regime may allow to overcome the heat transfer thresholds observed during the nucleate boiling crisis and offers a new pathway for heat transfer under microgravity conditions.
\end{abstract}

\maketitle

{\em Introduction}.-- The thermal energy transfer from a heater into a liquid is greatly increased once the boiling temperature is reached and vapor bubbles are formed. Then the previous convective driven flow becomes advected by the growing and detaching bubbles rising into the bulk under the action of gravity. By increasing the temperature of the heater further more bubbles are nucleated until a continuous vapor layer is formed~\cite{Dhir1998}. In this so-called film boiling regime heat transfer is strongly reduced; this regime limits the design and efficiency of common heaters. To overcome this boiling crisis, research is focused on the enhancement of heat transfer while avoiding the transition to film boiling. Current promising approaches are the texturing of the heater surface~\cite{Varanasi2015} or the use of thin electrically heated wires~\cite{PengBook}. Here we report a nucleate boiling regime on a flat substrate where the vapor bubble does not detach from the surface yet transports heat through a flow caused by bubble oscillations and thermocapillary stresses. This letter starts with a description of this unexpected oscillate boiling regime and relates the phenomenon with reports from literature having some similarity before we provide two simple models. The first model, based on a set of ordinary differential equations (ODEs), explains the orgin of the bubble oscillation and the dependency of the oscillation frequency on the heater power. The second model describes the thermal gradients and the resulting thermocapillary flow using a Navier Stokes solver which is compared to the experimental measurements. This novel oscillate boiling regime may allow to overcome the film boiling regime through an auto-oscillatory flow from a constant heat input. 

\begin{figure}
\includegraphics[width=\columnwidth]{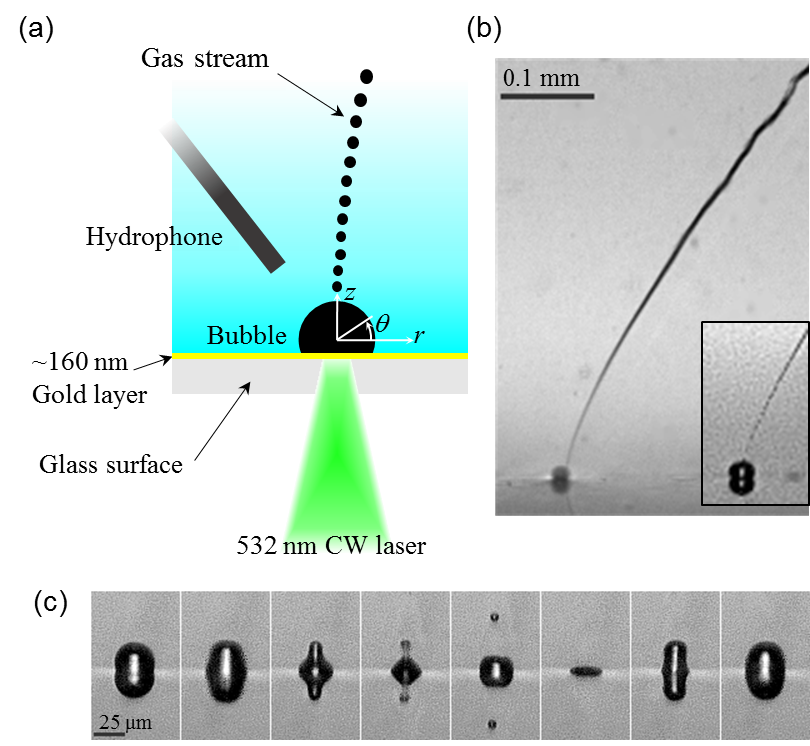}
\caption{Laser powered microheater: (a) A 532\,nm CW laser (29.5\,mW) is focused onto a thin layer of gold (around 165 nm thickness). (b) A bubble forms on the gold layer and sheds of microbubbles into the bulk (see inset). (c) Details of the bubble oscillation leading to the microbubble pinch-off at the bubble's apex.\label{fig1}}
\end{figure}

 {\em Experiment}.-- The heater is a glass plate with gold deposited onto it that is illuminated with a continuous wave focused laser as sketched in Fig.~1a. The power of the laser ($\lambda=532\,$nm, $1\,$W, Becen Optoelectronics, Shenzhen, China) is adjusted between $29.5\,$mW and $106.1\,$mW, the thickness of the gold layer is $165\,$nm, and the diameter of the laser focus is around $10\,\mu$m. After opening the laser shutter we observe a rapid vaporous cavitation event where one bubble is explosively expanding to a few hundred micrometer in diameter, rapidly shrinking, fragmenting, and dissolving. Yet a small bubble remains at the laser focus and grows in size and oscillates while its triple contact line remains pinned. The diameter of the contact area is around $15\,\mu$m and its maximum height during oscillation is about $13\,\mu$m. Interestingly, the unaided eye observes a dark streak on top of the bubble resembling a smoke screen as shown in Fig.~1b; yet using shorter exposure time it becomes evident that the dark screen is composed of microbubbles leaving the apex of the surface attached oscillating bubble as shown in the inset of Fig.~1b. 

Selected frames from a high-speed recording taken at short exposure times reveal the bubble oscillation in greater details, see Fig.~1c. The bubble expands into a spherical cap, reaches a maximum volume and shrinks with the contact line remaining pinned. This leads to a microbubble pinch-off from the bubble's apex while at minimum volume the bubble obtains a pancake shape. The bubble then grows back into an elongated shape before it reaches again an approximately spherical cap shape. For clarity we show in Fig.~1c the sequence of events during one oscillation period for a relatively large bubble of $R_{max}=21\mu$m. A similar sequence for a smaller bubble is shown in Ref.~\footnote{See supplementary material Fig. S2 at ... }. The stream of microbubbles in the liquid observed in Fig.~1b are formed by a steady bubble pinch-off from a pinned bubble oscillating near a rigid boundary. The rise velocities of the microbubbles close to the oscillating bubble are of the order of $2\,$cm/s while further away they decrease to $5.5\,$mm/s, indicating a background flow generated by the  oscillating bubble. 

\begin{figure}
\includegraphics[width=0.75\columnwidth]{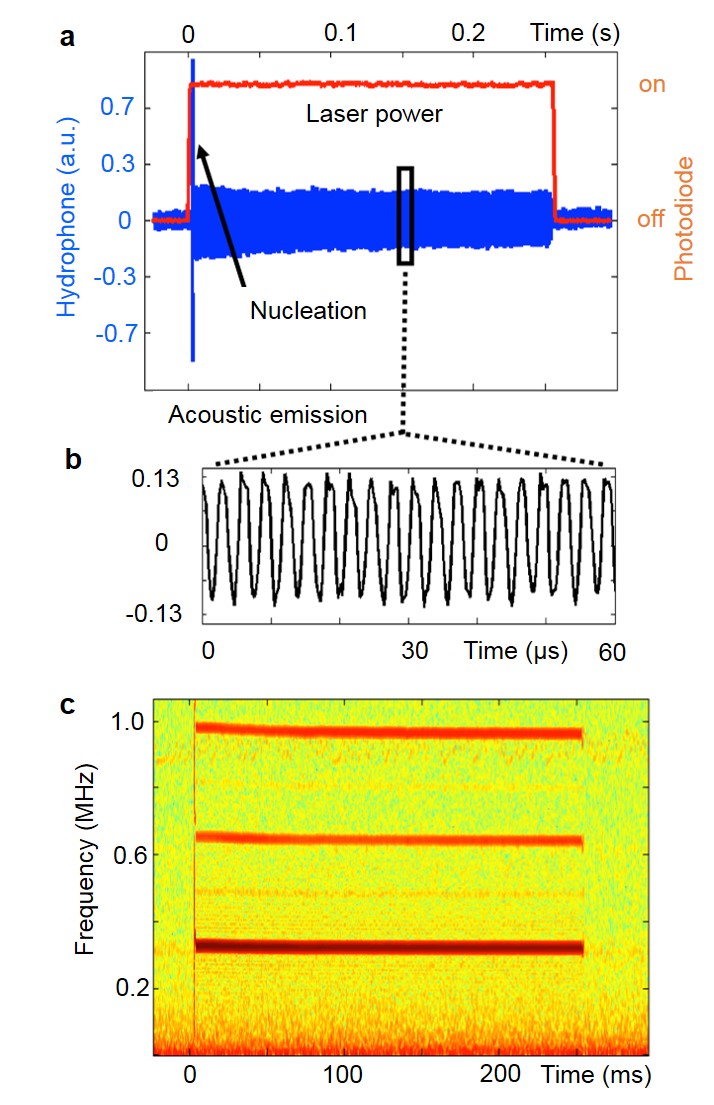}
\caption{Measurement of the bubble oscillation through acoustic emission. (a) The signal picked up by a hydrophone at nucleation and during continuous illumination with the laser beam. (b) Zoomed in region showing the sinusoidal signal oscillating with $320\,$kHz. (c) The spectrogram depicts the fundamental frequency and higher harmonics at nearly constant frequency over the duration of the laser heating of $250\,$ms.\label{fig2}}
\end{figure}

The clock-like repeatability of the bubble oscillation becomes evident from the recording of the acoustic emission. Signals from a hydrophone located close to the bubble (see Fig.~1a) are presented in Fig.~2. Figure~2a correlates the acoustic emission with the laser power monitored with a photodiode. After opening the shutter at time $t=0$ a large hydrophone signal is obtained as caused by the vaporous explosion; it is followed by a smaller and steady signal which stops at the moment the laser is switched off. Zooming into the hydrophone signal, Fig.~2b, we find a sinusoidal signal at a frequency of about $320\,$kHz emitted from the bubble. The time dependent Fourier analysis, the so-called  spectrogram of the hydrophone signal is shown in Fig.~2c. It reveals that after a brief broadband signal at $t \approx 0$~\footnote{It takes about $1\,$ms for the cavitation event to occur at the laser power of $29.5\,$mW.}, a strong fundamental frequency is observed which lasts until the laser is switched off at $t=250\,$ms.  
Figure~2c also reveals besides the fundamental frequency sharp bands of higher harmonics, which are expected for large amplitude bubble oscillations~\cite{LauterbornKurz2010}. 
% * <nguyenda001@e.ntu.edu.sg> 22:05:38 31 Jan 2016 UTC+0800:
% The oscillation can sustain for more than 20s given that laser illumination is provided. Writing this way gives the impression that after 250ms of illumination, a 20s oscillation can be achieved
% ^ <cdohl@ntu.edu.sg> 10:55:51 06 Apr 2016 UTC+0800:
% good, changes
The oscillation can be sustained for more than 20s (not shown here) given that laser illumination is provided. What stabilizes this mechanical oscillator for nearly $10$ million cycles while driven by a continuous power input and what sets the frequency?

Before we try to model this puzzling phenomenon we connect this experiment with literature: The first observations of oscillatory bubble dynamics from a continuous wave laser were reported by Sukhodolsky~\cite{Sukhodolsky1987,Sukhodolsky1990}. They termed the process thermocavitation to describe vapor bubble explosions created by a focused Argon Ion laser into a strongly light absorbing liquid. The bubbles collapse and detach from the surface before the cycle repeats. They report frequencies of several hundred cycles per second. In a more recent work using an IR diode bubble oscillations up to 4 KHz were found~\cite{RamosGarcia2010}. The main difference in the experimental arrangement between these thermocavitation experiments and the present oscillate boiling is that for the latter heat is supplied to the surface, not absorbed in the liquid. The resulting bubble oscillation in the thermocavitation experiments consist of the vapourous explosion, condensation of the bubble back into the liquid, followed by some heating time to reach again superheat; thus resulting into relatively large bubbles oscillating at low frequencies, very much different from the continuous bubble oscillations reported here in the 100 kHz regime and above.

The formation of a jet-resembling "structure" above a boiling bubble has been found for sufficiently thin wires heated in a subcooled liquid~\cite{PengBook}. This so-called "jet flow phenomena" is caused by thermocapillary flow of a stationary or translating bubble generated on a thin wire driving a liquid flow away from the hot wire.  

Next, we explain the origin of oscillation by taking into account conservation of momentum and energy. From this we'll give account to the large scale flow caused partially by thermal gradients.

{\em Model for bubble oscillation}.--
The bubble dynamics is simplified by assuming a hemispherical bubble containing gas and vapor which is undergoing oscillations at a boundary with only a radial velocity. This neglects the pinned triple phase contact line, unsteady boundary layers, and bubble pinch off, yet includes the liquid inertia and the restoring forces from the bubble content. With this assumption the momentum conservation equation for the fluid can be formulated as a $2^\mathrm{nd}$ order, non-linear ODE. Here we choose the Keller-Miksis model which has proven sufficiently accurate for high Reynolds number oscillation of surface attached bubbles~\cite{BremondPRL2006}. The Reynolds number supports this approach, $Re=2 U R_{max}/\nu\approx 100$ where $U$ is the bubble wall velocity and $\nu$ is the kinematic viscosity. We combine the momentum equation for the fluid with an energy equation for the bubble interior  using the first law of thermal dynamics: $pdV+dU=dQ$, which includes the work done to the gas and vapor content of the bubble, the change in potential energy accounting for evaporation and condensation, and the supplied energy through the laser minus the heat transfer into the liquid by diffusion and the latent heat. To close the equations we need to approximate the thermal boundary layer using the approach of Toegel et al.~\cite{toegel2003}. The energy equation is coupled to the Keller-Miksis equation through the bubble pressure assuming an ideal gas law. The equations are detailed in Ref.~\footnote{See supplementary material Sec. IV at ...  for the derivation of the equations and parameters used.}. The low heat transfer coefficient of gas and vapour prevents the heating of the bubble content by the laser. Thus, as expected, we find that an initial displacement of the bubble from its equilibrium radius leads to a strongly damped oscillation because of viscous stresses and acoustic emissions unless energy is fed into the bubble oscillator.

What is the mechanism of energy transfer? The bubble shapes in Fig.~1c reveal that during the collapsing phase of the bubble oscillation, a jet is formed which impinges onto the heated surface. This contact now allows heat transfer by vaporization of the liquid. We model this process within the scope of a hemi-spherical bubble model as kicked-oscillator. This is accomplished by adding a heat source through the laser power whenever the bubble is smaller than a critical radius, $R_{crit}$. We chose arbitrarily $R_{crit}=1\,\mu$m as it is clear that $R_{crit}\ll R^{ss}_{max}$ (constant steady-state radius), and in experiments we find a similar minimum bubble size. The argument that a jet transports liquid towards the surface is supported with experiments and is detailed in Ref. ~\footnote{See supplementary material Sec. III at  ... .}.

Figure~3a depicts the dynamics of the bubble oscillator for typical experimental parameters of water at $T_l=40^\circ$\,C and two initial conditions. We find that both radius-time curves show the same final dynamics, i.e. an inertial-driven bubble oscillation with strong collapses and a constant steady-state radius,  $R^{ss}_{max}$. A bubble starting with a radius $R(t=0)>R^{ss}_{max}$ decreases its amplitude within a few cycles while a bubble starting a smaller radius $R(t=0)<R^{ss}_{max}$ increases its amplitude. This supports the robustness of the bubble oscillator observed in experiments. The inset of Fig.~3a is a phase-space plot $(U=dR/dt,R)$ for these two initial conditions; it clearly demonstrates that both trajectories eventually converge to the same limit cycle.

The simple ODE model allows now to study the resonance frequency of the system as a function of the laser power or $R^{ss}_{max}$. For this plot the laser power is varied between $3\,$mW and $120\,$mW and the resulting steady state maximum radius and the fundamental frequency of oscillation of the bubble (resonance frequency) is measured. It is well known that the oscillation period of an inertial bubble near a boundary is prolonged. For hemispherical bubbles an increase of $k=1.2$ has been measured~\cite{vogel,maksimov2005} and is multiplied to the calculated radius $R^{ss}_{max}$.

We compare the prediction of the model with the experiment by plotting simulations and experimental data of the resonance frequency as a function of the steady state $R^{ss}_{max}$. Also, we compare the predicted and measured $R^{ss}_{max}$ as a function of the laser energy. With increasing laser power we see a monotonically decreasing resonance frequency $420\,$kHz $\geq f_{res} \geq 200\,$kHz, all having period-one solutions of the non-linear oscillator. The predicted bubble radius $R^{ss}_{max}$ increases from  $8.7\,\mu$m to $20\,\mu$m. The model captures the experiments quantitatively for the smaller bubbles, while some difference is observed for largest bubbles which are about 30\% smaller than predicted. We also validate the model by comparing the measured laser power absorbed within the gold film with the power input of the source used in the simulations, see dashed curve in Fig.~3b. Experimentally, the absorbed power is obtained by measuring the reflected and the transmitted laser power and subtracting from the total laser power. Three laser powers have been tested with 3 experimental runs each. Again good agreement with the simple ODE model is achieved supporting that the important fluid and thermal dynamics is accounted for.  

\begin{figure}
\includegraphics[width=0.8\columnwidth]{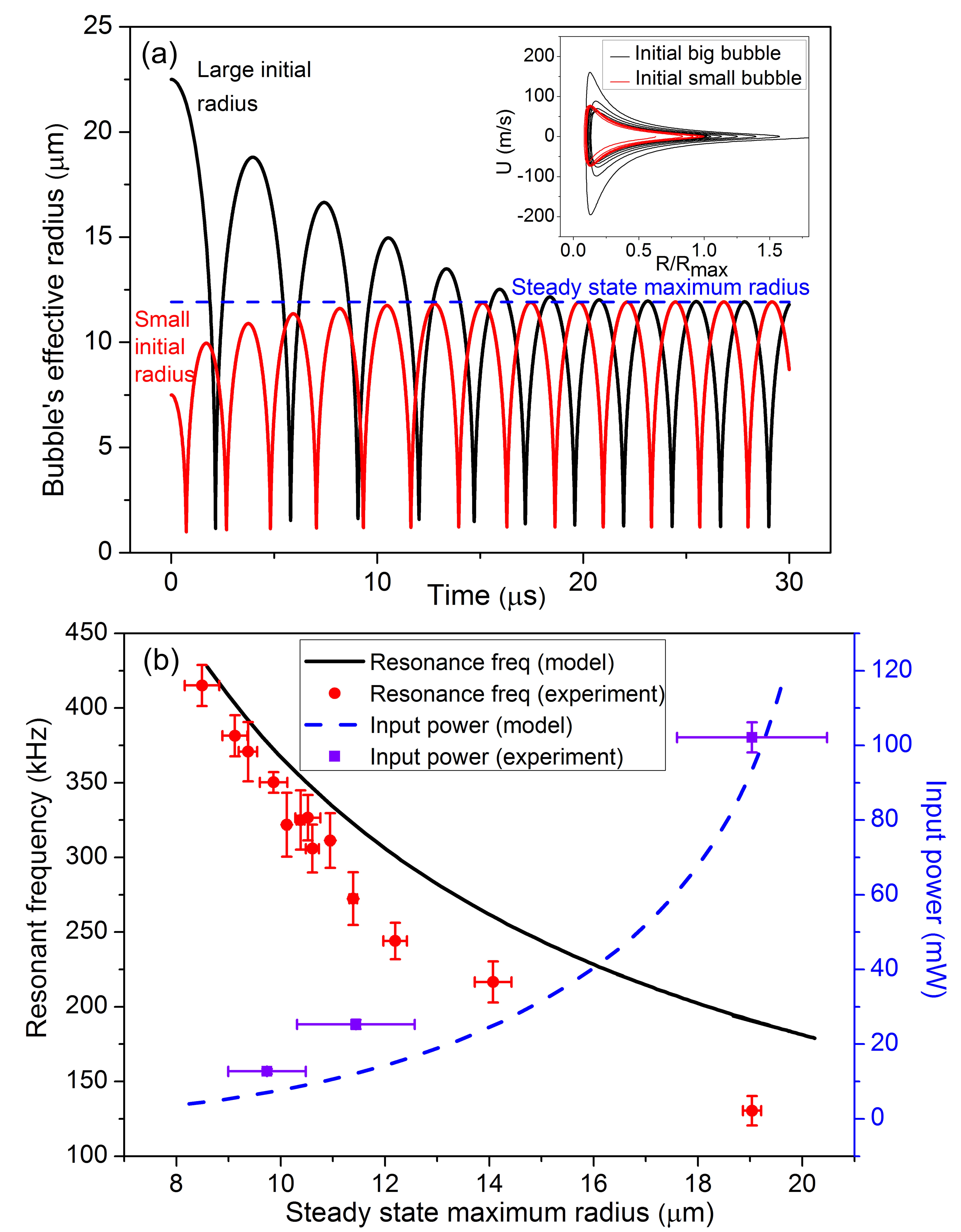}
\caption{a) Solution of the kicked bubble oscillator model (see text) for two initial conditions $R(t=0)$. One starting at a bubble radius larger than the steady state maximum radius $R^{ss}_{max}$ and one at a smaller radius. The phase-space inset demonstrates that both trajectories approach the same attractor. Bubble's effective radius is calculated by assuming a hemispherical cap of the same volume. b) Dependence of the resonance frequency (solid line) and laser power (dashed line) on the maximum bubble radius $R^{ss}_{max}$. The experimental data points are plotted as filled circles and squares.
    \label{fig3}}
\end{figure}

{\em Model for thermocapillary flow}.--  The small size of the heater suggests that large thermal gradients may lead to a thermocapillary flow. Therefore, we model the thermal field to obtain the Marangoni stresses on the bubble interface and the resulting flow pattern. As we have been measuring only the time-averaged flow we simplify the problem by neglecting the bubble oscillation; the bubble has then the time averaged shape of a spherical cap.

 The computational domain consists of two parts: the bubble domain and the liquid domain. The bubble domain has a constant temperature following Clausius-Clapeyron equation. For the liquid domains, the governing equations includes the Navier-Stokes, mass conservation, and energy transfer equation. At the gas-liquid interface, the thermocapillary stress term 
\begin{equation}
\tau_{R\theta}=-\frac{1}{R}\frac{d\sigma}{dT_{i}}\left(\frac{\partial T_{i}}{\partial\theta}\right)
\end{equation}
needs to be accounted for. The temperature dependent surface tension adds a stress component in the $(r,\theta)$ plane, see Eq. (1) and Fig.~1a. This thermocapillary stress accelerates the liquid from high temperature regions near the contact line to the bubble apex. The simulation results are shown in Fig.~4a with a bubble diameter of $17\mu$m and laser power of 40mW. The flow field profile is presented on the left and the temperature distribution is presented on the right. The flow field profile shows a Marangoni flow at the gas-liquid interface. This flow drives liquid upward at the top and toward the bubble at the side. The upward flow (around 40 cm/s close to the bubble's top surface, and decreasing further away) contributes to the rise of pinched-off bubble and significantly enhances heat transfer from the hot substate to the cold liquid in the upper region. Meanwhile, the flow at the side drives liquid towards the bubble and causes a cooling of the substrate, i.e. stabilises the thermal gradients near the bubble. Thereby, the boiling region remains localised. Although the present simulations don't account for this, we understand that the recirculation of cold liquid is essential for the stability of the bubble oscillation.
The general features of the flow field of Fig.~4a (left) are found in experiments using particle image velocimetry, Fig.~4b. There we see a central flow pointing upward and a cut in the $(r,z)$-plane through a vortex ring as predicted in the simulations. The maximum velocities are of the order of 10\,mm/s approximately $200\,\mu$m above the oscillating bubble while simulations predict a value of $2\,$mm/s at the same location. We speculate that this difference is caused by the neglected microbubble stream which through buoyancy accelerates the liquid even far from the oscillating bubble.

\begin{figure}
\includegraphics[width=\columnwidth]{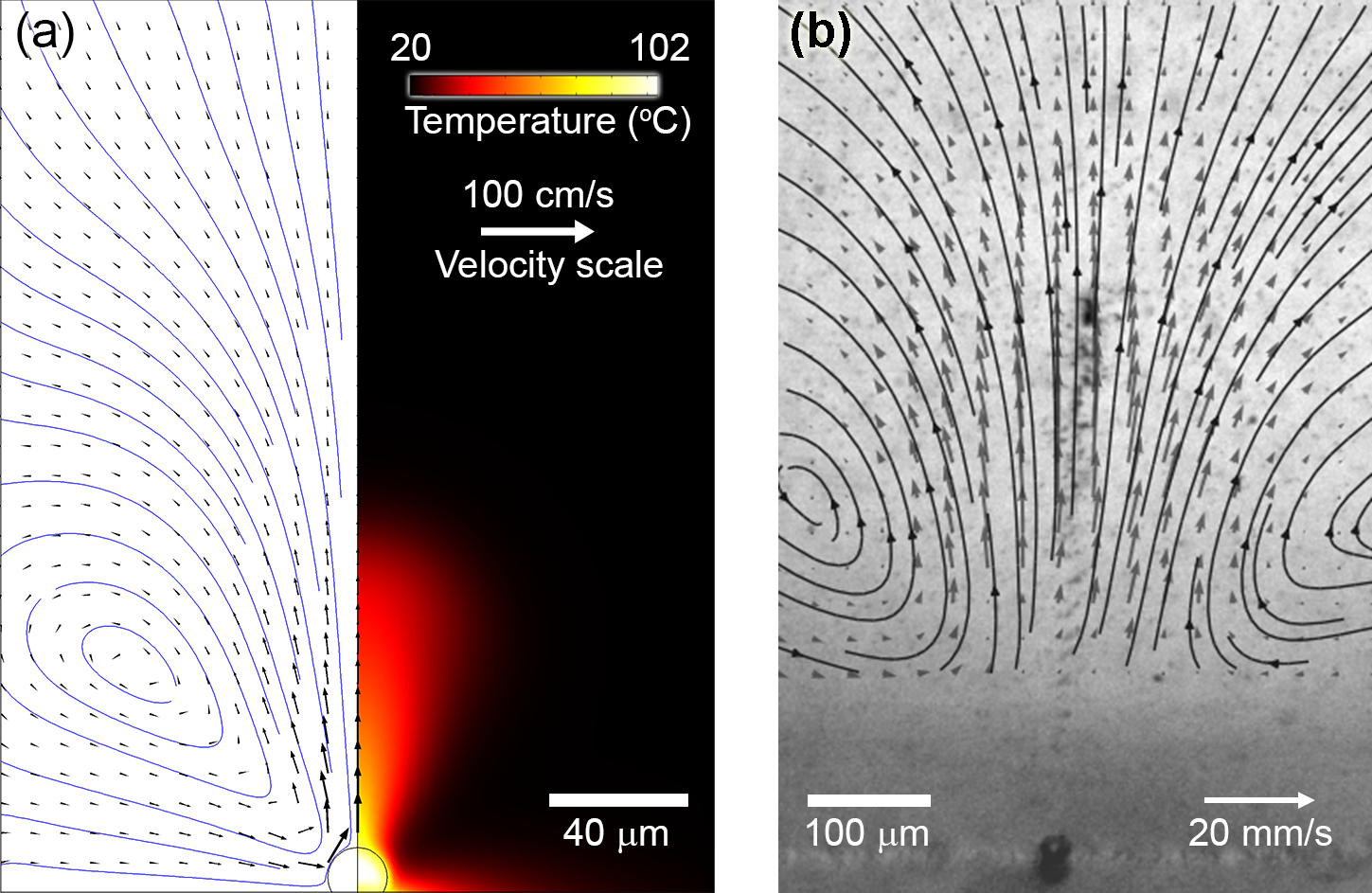}
\caption{a) Simulation of the heat transfer (right) and the thermocapillary induced flow (left) of a steady bubble under laser irradiation of 40 mW. b) Flow field obtained from particle image velocimetry with $2\,\mu$m diameter particles. Due to illumination constrains we could not identify particles in the bottom part of the image. Note the different image scales on the left and right figure.\label{fig4}}
\end{figure}
  
{\em Discussion}.--
We speculate that the critical ingredients of the oscillate boiling phenomenon are the stable pinning of the contact line together with the small size of the heater. The non-spherical collapse of the bubble transports liquid on the substrate where it initiates the re-expansion of the bubble. The heat transfer simulations demonstrate that the thermocapillary flow qualitatively explain the observed flow field. Besides thermocapillary flow, bubbles in contact with a surface and undergoing volume oscillations can generate a streaming flow and show a rather similar flow pattern~\cite{Hilgenfeld}. Here we have not investigated the importance of the acoustic streaming flow as the small amplitude approximation for the bubble oscillation is not valid for the inertial collapse. Pinch off of microbubbles has been previously observed for resonantly driven bubbles in hydrophobic pits~\cite{borkent2009, zijlstra2015}.  

A hot stream (jet) of liquid from the top of a bubble has also been reported for nucleate boiling on thin wires and termed {\em jet flow} phenomena. There, the bubbles are generally moving along the wire and reveal complex dynamics and interaction. For one particular case of these jet flows termed the "bubble bunch jet flow", it was speculated to consist of a stream of vapour bubbles~\cite{PengBook} (page 66). Interestingly, this  observation was made close to the critical heat flux that may relate to the loss of the liquid microlayer~\cite{zou2016} which resulted into a pinning of the three phase contact line. 

A second manifestation of the oscillate boiling regime may have been reported by   
MacDonald {\em et al.}~\cite{macdonald2003}. They inferred from a high frequency oscillations picked up with a photodiode that a bubble oscillates at the tip of an optical fiber submerged into liquid nitrogen. The bubble may have been generated with $1\,\mu$s long laser pulses absorbed at the tip of a metallic film-coated fiber ($9\,\mu$m core diameter). They observed bubble oscillations lasting for $20\,\mu$s with a decaying frequency starting from $17\,$MHz!

The observations of this new regime may offer opportunities for heat transfer applications. For example, in conventional boiling applications, gravity is essential to remove the bubbles from the surface, while in the oscillate boiling regime, heat is transported by thermocapillarity, thus in the direction from hot to cold independent of buoyancy. This qualifies this regime for microgravity environments. Another interesting aspect of the regime is that the bubble is generated above the Leidenfrost temperature~\cite{quere2013}. By using multiple heaters it may be possible to increase the heat transfer while avoiding the boiling crisis. We suggest experiments where the effect of neighbouring bubbles should be studied to determine the closest packing of bubbles while retaining the oscillate boiling regime.
 
{\em Acknowledgements}.--
The research is supported by MOE Singapore (Tier 1 grant RG90/15).
We thank Hongjie An and Tran Anh Tuan for their help with the experiments and David Qu\'{e}r\'{e} and Andrea Prosperetti for enlightening discussions.

\bibliography{oscillate_boiling} % Tell bibtex which .bib file to use

\begin{thebibliography}{21}%
\makeatletter
\providecommand \@ifxundefined [1]{%
 \@ifx{#1\undefined}
}%
\providecommand \@ifnum [1]{%
 \ifnum #1\expandafter \@firstoftwo
 \else \expandafter \@secondoftwo
 \fi
}%
\providecommand \@ifx [1]{%
 \ifx #1\expandafter \@firstoftwo
 \else \expandafter \@secondoftwo
 \fi
}%
\providecommand \natexlab [1]{#1}%
\providecommand \enquote  [1]{``#1''}%
\providecommand \bibnamefont  [1]{#1}%
\providecommand \bibfnamefont [1]{#1}%
\providecommand \citenamefont [1]{#1}%
\providecommand \href@noop [0]{\@secondoftwo}%
\providecommand \href [0]{\begingroup \@sanitize@url \@href}%
\providecommand \@href[1]{\@@startlink{#1}\@@href}%
\providecommand \@@href[1]{\endgroup#1\@@endlink}%
\providecommand \@sanitize@url [0]{\catcode `\\12\catcode `\$12\catcode
  `\&12\catcode `\#12\catcode `\^12\catcode `\_12\catcode `\%12\relax}%
\providecommand \@@startlink[1]{}%
\providecommand \@@endlink[0]{}%
\providecommand \url  [0]{\begingroup\@sanitize@url \@url }%
\providecommand \@url [1]{\endgroup\@href {#1}{\urlprefix }}%
\providecommand \urlprefix  [0]{URL }%
\providecommand \Eprint [0]{\href }%
\providecommand \doibase [0]{http://dx.doi.org/}%
\providecommand \selectlanguage [0]{\@gobble}%
\providecommand \bibinfo  [0]{\@secondoftwo}%
\providecommand \bibfield  [0]{\@secondoftwo}%
\providecommand \translation [1]{[#1]}%
\providecommand \BibitemOpen [0]{}%
\providecommand \bibitemStop [0]{}%
\providecommand \bibitemNoStop [0]{.\EOS\space}%
\providecommand \EOS [0]{\spacefactor3000\relax}%
\providecommand \BibitemShut  [1]{\csname bibitem#1\endcsname}%
\let\auto@bib@innerbib\@empty
%</preamble>
\bibitem [{\citenamefont {Dhir}(1998)}]{Dhir1998}%
  \BibitemOpen
  \bibfield  {author} {\bibinfo {author} {\bibfnamefont {V.}~\bibnamefont
  {Dhir}},\ }\href {\doibase {10.1146/annurev.fluid.30.1.365}} {\bibfield
  {journal} {\bibinfo  {journal} {{Annu. Rev. Fluid Mech.}}\ }\textbf {\bibinfo
  {volume} {{30}}},\ \bibinfo {pages} {365} (\bibinfo {year}
  {{1998}})}\BibitemShut {NoStop}%
\bibitem [{\citenamefont {Dhillon}\ \emph {et~al.}(2015)\citenamefont
  {Dhillon}, \citenamefont {Buongiorno},\ and\ \citenamefont
  {Varanasi}}]{Varanasi2015}%
  \BibitemOpen
  \bibfield  {author} {\bibinfo {author} {\bibfnamefont {N.~S.}\ \bibnamefont
  {Dhillon}}, \bibinfo {author} {\bibfnamefont {J.}~\bibnamefont {Buongiorno}},
  \ and\ \bibinfo {author} {\bibfnamefont {K.~K.}\ \bibnamefont {Varanasi}},\
  }\href@noop {} {\bibfield  {journal} {\bibinfo  {journal} {{Nature Commun.}}\
  }\textbf {\bibinfo {volume} {{6}}} (\bibinfo {year} {{2015}})}\BibitemShut
  {NoStop}%
\bibitem [{\citenamefont {Peng}(2011)}]{PengBook}%
  \BibitemOpen
  \bibfield  {author} {\bibinfo {author} {\bibfnamefont {X.}~\bibnamefont
  {Peng}},\ }\enquote {\bibinfo {title} {{Micro Transport Phenomena During
  Boiling}},}\ \ (\bibinfo  {publisher} {Springer Science \& Business Media},\
  \bibinfo {year} {{2011}})\ pp.\ \bibinfo {pages} {{1--255}}\BibitemShut
  {NoStop}%
\bibitem [{Note1()}]{Note1}%
  \BibitemOpen
  \bibinfo {note} {See supplementary material Fig. S2 at ...}\BibitemShut
  {Stop}%
\bibitem [{Note2()}]{Note2}%
  \BibitemOpen
  \bibinfo {note} {It takes about $1\protect \,$ms for the cavitation event to
  occur at the laser power of $29.5\protect \,$mW.}\BibitemShut {Stop}%
\bibitem [{\citenamefont {Lauterborn}\ and\ \citenamefont
  {Kurz}(2010)}]{LauterbornKurz2010}%
  \BibitemOpen
  \bibfield  {author} {\bibinfo {author} {\bibfnamefont {W.}~\bibnamefont
  {Lauterborn}}\ and\ \bibinfo {author} {\bibfnamefont {T.}~\bibnamefont
  {Kurz}},\ }\href@noop {} {\bibfield  {journal} {\bibinfo  {journal} {Rep.
  Prog. Phys.}\ }\textbf {\bibinfo {volume} {73}},\ \bibinfo {pages} {106501}
  (\bibinfo {year} {2010})}\BibitemShut {NoStop}%
\bibitem [{\citenamefont {Rastopov}\ and\ \citenamefont
  {Sukhodolsky}(1987)}]{Sukhodolsky1987}%
  \BibitemOpen
  \bibfield  {author} {\bibinfo {author} {\bibfnamefont {S.}~\bibnamefont
  {Rastopov}}\ and\ \bibinfo {author} {\bibfnamefont {A.}~\bibnamefont
  {Sukhodolsky}},\ }\href@noop {} {\bibfield  {journal} {\bibinfo  {journal}
  {{Dokl. Akad. Nauk SSSR}}\ }\textbf {\bibinfo {volume} {{295}}},\ \bibinfo
  {pages} {1108} (\bibinfo {year} {{1987}})}\BibitemShut {NoStop}%
\bibitem [{\citenamefont {Rastopov}\ and\ \citenamefont
  {Sukhodolsky}(1991)}]{Sukhodolsky1990}%
  \BibitemOpen
  \bibfield  {author} {\bibinfo {author} {\bibfnamefont {S.}~\bibnamefont
  {Rastopov}}\ and\ \bibinfo {author} {\bibfnamefont {A.}~\bibnamefont
  {Sukhodolsky}},\ }in\ \href@noop {} {\emph {\bibinfo {booktitle} {Proceedings
  Of The Society Of Photo-Optical Instrumentation Engineers (SPIE)}}},\ Vol.\
  \bibinfo {volume} {{1440}}\ (\bibinfo {organization} {{International Society
  for Optics and Photonics}},\ \bibinfo {year} {{1991}})\ pp.\ \bibinfo {pages}
  {127--134}\BibitemShut {NoStop}%
\bibitem [{\citenamefont {Ramirez-San-Juan}\ \emph {et~al.}(2010)\citenamefont
  {Ramirez-San-Juan}, \citenamefont {Rodriguez-Aboytes}, \citenamefont
  {Martinez-Canton}, \citenamefont {Baldovino-Pantaleon}, \citenamefont
  {Robledo-Martinez}, \citenamefont {Korneev},\ and\ \citenamefont
  {Ramos-Garcia}}]{RamosGarcia2010}%
  \BibitemOpen
  \bibfield  {author} {\bibinfo {author} {\bibfnamefont {J.~C.}\ \bibnamefont
  {Ramirez-San-Juan}}, \bibinfo {author} {\bibfnamefont {E.}~\bibnamefont
  {Rodriguez-Aboytes}}, \bibinfo {author} {\bibfnamefont {A.~E.}\ \bibnamefont
  {Martinez-Canton}}, \bibinfo {author} {\bibfnamefont {O.}~\bibnamefont
  {Baldovino-Pantaleon}}, \bibinfo {author} {\bibfnamefont {A.}~\bibnamefont
  {Robledo-Martinez}}, \bibinfo {author} {\bibfnamefont {N.}~\bibnamefont
  {Korneev}}, \ and\ \bibinfo {author} {\bibfnamefont {R.}~\bibnamefont
  {Ramos-Garcia}},\ }\href {\doibase {10.1364/OE.18.008735}} {\bibfield
  {journal} {\bibinfo  {journal} {{Opt. Express}}\ }\textbf {\bibinfo {volume}
  {{18}}},\ \bibinfo {pages} {8735} (\bibinfo {year} {{2010}})}\BibitemShut
  {NoStop}%
\bibitem [{\citenamefont {Bremond}\ \emph {et~al.}(2006)\citenamefont
  {Bremond}, \citenamefont {Arora}, \citenamefont {Ohl},\ and\ \citenamefont
  {Lohse}}]{BremondPRL2006}%
  \BibitemOpen
  \bibfield  {author} {\bibinfo {author} {\bibfnamefont {N.}~\bibnamefont
  {Bremond}}, \bibinfo {author} {\bibfnamefont {M.}~\bibnamefont {Arora}},
  \bibinfo {author} {\bibfnamefont {C.-D.}\ \bibnamefont {Ohl}}, \ and\
  \bibinfo {author} {\bibfnamefont {D.}~\bibnamefont {Lohse}},\ }\href@noop {}
  {\bibfield  {journal} {\bibinfo  {journal} {Phy. Rev. Lett.}\ }\textbf
  {\bibinfo {volume} {96}},\ \bibinfo {pages} {224501} (\bibinfo {year}
  {2006})}\BibitemShut {NoStop}%
\bibitem [{\citenamefont {Toegel}\ and\ \citenamefont
  {Lohse}(2003)}]{toegel2003}%
  \BibitemOpen
  \bibfield  {author} {\bibinfo {author} {\bibfnamefont {R.}~\bibnamefont
  {Toegel}}\ and\ \bibinfo {author} {\bibfnamefont {D.}~\bibnamefont {Lohse}},\
  }\href@noop {} {\bibfield  {journal} {\bibinfo  {journal} {J. Chem. Phys.}\
  }\textbf {\bibinfo {volume} {118}},\ \bibinfo {pages} {1863} (\bibinfo {year}
  {2003})}\BibitemShut {NoStop}%
\bibitem [{Note3()}]{Note3}%
  \BibitemOpen
  \bibinfo {note} {See supplementary material Sec. IV at ... for the derivation
  of the equations and parameters used.}\BibitemShut {Stop}%
\bibitem [{Note4()}]{Note4}%
  \BibitemOpen
  \bibinfo {note} {See supplementary material Sec. III at ... .}\BibitemShut
  {Stop}%
\bibitem [{\citenamefont {Godwin}\ \emph {et~al.}(1999)\citenamefont {Godwin},
  \citenamefont {Chapyak}, \citenamefont {Noack},\ and\ \citenamefont
  {Vogel}}]{vogel}%
  \BibitemOpen
  \bibfield  {author} {\bibinfo {author} {\bibfnamefont {R.~P.}\ \bibnamefont
  {Godwin}}, \bibinfo {author} {\bibfnamefont {E.~J.}\ \bibnamefont {Chapyak}},
  \bibinfo {author} {\bibfnamefont {J.}~\bibnamefont {Noack}}, \ and\ \bibinfo
  {author} {\bibfnamefont {A.}~\bibnamefont {Vogel}},\ }in\ \href@noop {}
  {\emph {\bibinfo {booktitle} {BiOS'99 International Biomedical Optics
  Symposium}}}\ (\bibinfo {organization} {International Society for Optics and
  Photonics},\ \bibinfo {year} {1999})\ pp.\ \bibinfo {pages}
  {225--236}\BibitemShut {NoStop}%
\bibitem [{\citenamefont {Maksimov}(2005)}]{maksimov2005}%
  \BibitemOpen
  \bibfield  {author} {\bibinfo {author} {\bibfnamefont {A.}~\bibnamefont
  {Maksimov}},\ }\href {\doibase {10.1016/j.jsv.2004.05.021}} {\bibfield
  {journal} {\bibinfo  {journal} {{J. Sound Vibr.}}\ }\textbf {\bibinfo
  {volume} {{283}}},\ \bibinfo {pages} {915} (\bibinfo {year}
  {{2005}})}\BibitemShut {NoStop}%
\bibitem [{\citenamefont {Marmottant}\ and\ \citenamefont
  {Hilgenfeldt}(2003)}]{Hilgenfeld}%
  \BibitemOpen
  \bibfield  {author} {\bibinfo {author} {\bibfnamefont {P.}~\bibnamefont
  {Marmottant}}\ and\ \bibinfo {author} {\bibfnamefont {S.}~\bibnamefont
  {Hilgenfeldt}},\ }\href@noop {} {\bibfield  {journal} {\bibinfo  {journal}
  {Nature}\ }\textbf {\bibinfo {volume} {423}},\ \bibinfo {pages} {153}
  (\bibinfo {year} {2003})}\BibitemShut {NoStop}%
\bibitem [{\citenamefont {Borkent}\ \emph {et~al.}(2009)\citenamefont
  {Borkent}, \citenamefont {Gekle}, \citenamefont {Prosperetti},\ and\
  \citenamefont {Lohse}}]{borkent2009}%
  \BibitemOpen
  \bibfield  {author} {\bibinfo {author} {\bibfnamefont {B.~M.}\ \bibnamefont
  {Borkent}}, \bibinfo {author} {\bibfnamefont {S.}~\bibnamefont {Gekle}},
  \bibinfo {author} {\bibfnamefont {A.}~\bibnamefont {Prosperetti}}, \ and\
  \bibinfo {author} {\bibfnamefont {D.}~\bibnamefont {Lohse}},\ }\href@noop {}
  {\bibfield  {journal} {\bibinfo  {journal} {Physics of Fluids
  (1994-present)}\ }\textbf {\bibinfo {volume} {21}},\ \bibinfo {pages}
  {102003} (\bibinfo {year} {2009})}\BibitemShut {NoStop}%
\bibitem [{\citenamefont {Zijlstra}\ \emph {et~al.}(2015)\citenamefont
  {Zijlstra}, \citenamefont {Fernandez~Rivas}, \citenamefont {Gardeniers},
  \citenamefont {Versluis},\ and\ \citenamefont {Lohse}}]{zijlstra2015}%
  \BibitemOpen
  \bibfield  {author} {\bibinfo {author} {\bibfnamefont {A.}~\bibnamefont
  {Zijlstra}}, \bibinfo {author} {\bibfnamefont {D.}~\bibnamefont
  {Fernandez~Rivas}}, \bibinfo {author} {\bibfnamefont {H.~J.}\ \bibnamefont
  {Gardeniers}}, \bibinfo {author} {\bibfnamefont {M.}~\bibnamefont
  {Versluis}}, \ and\ \bibinfo {author} {\bibfnamefont {D.}~\bibnamefont
  {Lohse}},\ }\href {\doibase 10.1016/j.ultras.2014.10.002} {\bibfield
  {journal} {\bibinfo  {journal} {Ultrasonics}\ }\textbf {\bibinfo {volume}
  {56}},\ \bibinfo {pages} {512} (\bibinfo {year} {2015})}\BibitemShut
  {NoStop}%
\bibitem [{\citenamefont {Zou}\ \emph {et~al.}(2016)\citenamefont {Zou},
  \citenamefont {Chanana}, \citenamefont {Agrawal}, \citenamefont {Wayner},\
  and\ \citenamefont {Maroo}}]{zou2016}%
  \BibitemOpen
  \bibfield  {author} {\bibinfo {author} {\bibfnamefont {A.}~\bibnamefont
  {Zou}}, \bibinfo {author} {\bibfnamefont {A.}~\bibnamefont {Chanana}},
  \bibinfo {author} {\bibfnamefont {A.}~\bibnamefont {Agrawal}}, \bibinfo
  {author} {\bibfnamefont {P.~C.}\ \bibnamefont {Wayner}}, \ and\ \bibinfo
  {author} {\bibfnamefont {S.~C.}\ \bibnamefont {Maroo}},\ }\href
  {http://dx.doi.org/10.1038/srep20240} {\bibfield  {journal} {\bibinfo
  {journal} {Sci. Rep.}\ }\textbf {\bibinfo {volume} {6}},\ \bibinfo {pages}
  {20240 EP } (\bibinfo {year} {2016})}\BibitemShut {NoStop}%
\bibitem [{\citenamefont {MacDonald}\ \emph {et~al.}(2003)\citenamefont
  {MacDonald}, \citenamefont {Fedotov}, \citenamefont {Pochon}, \citenamefont
  {Soares}, \citenamefont {Zheludev}, \citenamefont {Guignard}, \citenamefont
  {Mihaescu},\ and\ \citenamefont {Besnard}}]{macdonald2003}%
  \BibitemOpen
  \bibfield  {author} {\bibinfo {author} {\bibfnamefont {K.}~\bibnamefont
  {MacDonald}}, \bibinfo {author} {\bibfnamefont {V.}~\bibnamefont {Fedotov}},
  \bibinfo {author} {\bibfnamefont {S.}~\bibnamefont {Pochon}}, \bibinfo
  {author} {\bibfnamefont {B.}~\bibnamefont {Soares}}, \bibinfo {author}
  {\bibfnamefont {N.}~\bibnamefont {Zheludev}}, \bibinfo {author}
  {\bibfnamefont {C.}~\bibnamefont {Guignard}}, \bibinfo {author}
  {\bibfnamefont {A.}~\bibnamefont {Mihaescu}}, \ and\ \bibinfo {author}
  {\bibfnamefont {P.}~\bibnamefont {Besnard}},\ }\href@noop {} {\bibfield
  {journal} {\bibinfo  {journal} {{Phy. Rev. E}}\ }\textbf {\bibinfo {volume}
  {{68}}} (\bibinfo {year} {{2003}})}\BibitemShut {NoStop}%
\bibitem [{\citenamefont {Qu\'{e}r\'{e}}(2013)}]{quere2013}%
  \BibitemOpen
  \bibfield  {author} {\bibinfo {author} {\bibfnamefont {D.}~\bibnamefont
  {Qu\'{e}r\'{e}}},\ }\href {\doibase 10.1146/annurev-fluid-011212-140709}
  {\bibfield  {journal} {\bibinfo  {journal} {Ann. Rev. Fluid Mech.}\ }\textbf
  {\bibinfo {volume} {45}},\ \bibinfo {pages} {197} (\bibinfo {year}
  {2013})}\BibitemShut {NoStop}%
\end{thebibliography}%

\end{document}